\documentclass[twocolumn,showpacs,preprintnumbers,nofootinbib,prd,
superscriptaddress,10pt]{revtex4-2}

\pdfoutput=1

\usepackage{amsmath,amssymb}
\usepackage{amsfonts}
\usepackage[normalem]{ulem}
\usepackage{textcomp}
\usepackage{hyperref}
\usepackage{enumitem}
\usepackage{bm}
\usepackage{afterpage}
\usepackage{graphicx}
\graphicspath{{img/}} 
\usepackage{psfrag}
\usepackage{mathtools}
\usepackage{tensor}
\usepackage{layouts}
\usepackage{DejaVuSans}
\usepackage{epstopdf}
\usepackage[usenames,dvipsnames]{xcolor}
\usepackage[utf8]{inputenc}
\usepackage{multirow}
\usepackage{rotating}
\usepackage{tabularx}
\usepackage{ragged2e}
\usepackage{blindtext}
\usepackage{orcidlink}
\newcommand{\R}{\mathbb{R}}

\DeclareMathOperator*{\argmin}{arg\,min}

\newcommand{\rescalar}[2]{( #1 |#2 )}

\begin{document}

\begin{abstract}
We introduce a machine learning model designed to rapidly and accurately predict the time domain gravitational wave emission of non-precessing binary black hole coalescences, incorporating the effects of higher order modes of the multipole expansion of the waveform. 
Expanding on our prior work \cite{Schmidt:2020yuu}, we decompose each mode by amplitude and phase and reduce dimensionality using principal component analysis. An ensemble of artificial neural networks is trained to learn the relationship between orbital parameters and the low-dimensional representation of each mode.
Our model is trained with $\sim 10^5$ signals with mass ratio $q \in [1,10]$ and dimensionless spins $\chi_i \in [-0.9, 0.9]$, generated with the state-of-the-art approximant \texttt{SEOBNRv4HM}, and it is able to generate waveforms up to $\sim 4\times 10^5 M$ long. We find that it achieves a median faithfulness of $10^{-4}$ averaged across the parameter space. We show that our model generates a single waveform two orders of magnitude faster than the training model, with the speed up increasing when waveforms are generated in batches.
This framework is entirely general and can be applied to any other time domain approximant capable of generating waveforms from aligned spin circular binaries, possibly incorporating higher order modes.
\end{abstract}
	
	\title{Generating Higher Order Modes from Binary Black Hole mergers with Machine Learning}
	\author{Tim \surname{Grimbergen}}
        \affiliation{Institute for Gravitational and Subatomic Physics (GRASP),
		Utrecht University, Princetonplein 1, 3584 CC Utrecht, The Netherlands}
	\author{Stefano \surname{Schmidt} \orcidlink{0000-0002-8206-8089}}
		\email{s.schmidt@uu.nl}
        \affiliation{Institute for Gravitational and Subatomic Physics (GRASP),
		Utrecht University, Princetonplein 1, 3584 CC Utrecht, The Netherlands}
        \affiliation{Nikhef, Science Park 105, 1098 XG, Amsterdam, The Netherlands}
	\author{Chinmay \surname{Kalaghatgi} \orcidlink{0000-0002-4688-867X}}
        \affiliation{Institute for Gravitational and Subatomic Physics (GRASP),
		Utrecht University, Princetonplein 1, 3584 CC Utrecht, The Netherlands}
        \affiliation{Nikhef, Science Park 105, 1098 XG, Amsterdam, The Netherlands}
	\author{Chris \surname{van den Broeck} \orcidlink{0000-0001-6800-4006}}
        \affiliation{Institute for Gravitational and Subatomic Physics (GRASP),
		Utrecht University, Princetonplein 1, 3584 CC Utrecht, The Netherlands}
        \affiliation{Nikhef, Science Park 105, 1098 XG, Amsterdam, The Netherlands}
	\maketitle

\section{Introduction}
\label{sec:intro}

With almost a hundred of confirmed detections, gravitational wave (GW) astronomy is entering a mature state, 
where many loud GW events will force the scientific community to develop faster analyses to deliver 
precision measurements. Expanding on past results \cite{LIGOScientific:2018mvr, LIGOScientific:2020ibl, LIGOScientific:2021usb}, 
the recent transient catalogue GWTC-3 \cite{KAGRA:2021vkt} is the latest achievment of the effort carried on by the 
LIGO-Virgo-KAGRA collaboration \cite{KAGRA:2013rdx, LIGOScientific:2014pky, VIRGO:2014yos, KAGRA:2020tym} 
and it relies on both instrument and data analysis development.

A crucial element of the data analysis is the ability to quickly and accurately generate waveforms for GW signals 
emitted by coalescing binary black holes (BBHs). Such waveforms are used for the expensive Bayesian estimation of 
the parameters characterizing a BBH \cite{Veitch:2014wba}: the analysis of a single event requires the online 
generation of up to billions of waveforms. As we move towards the next generation of detectors, such as Einstein Telescope 
\cite{Punturo:2010zz, Maggiore:2019uih} and Cosmic Explorer \cite{Reitze:2019iox, Evans:2021gyd}, it will become paramount to deploy accurate waveform models 
that are fast and, at the same time, incorporate the full physics of the problem, otherwise our analyses will  
become subject to systematic errors in the parameter recovery \cite{Purrer:2019jcp}. On the other hand, 
accurate models are often slow to generate on a computer and the analyses might struggle to keep up with the large 
event rate expected in the next-generation observatories  \cite{Samajdar:2021egv}.
Balancing the two needs is challenging, since speed and accuracy are often at trade.

An essential aspect for a realistic BBH signal model is the incorporation of higher-order modes (HMs) of the
multipole expansion of the waveform \cite{Maggiore:2007ulw}. For nearly equal mass systems, 
the leading-order mode is orders of magnitude larger than the others and, including the HMs does not 
significantly affect the parameter estimation. However, it has been demonstrated \cite{Varma:2014jxa, Varma:2016dnf, Roy:2019phx, Mills:2020thr} 
that HMs are observable in highly asymmetric binary systems. In fact, the effect of HMs has already been observed 
in at least two BBH events originating from asymmetric binaries \cite{LIGOScientific:2020stg, LIGOScientific:2020zkf}. This underscores the importance of including HMs in any parameter estimation pipeline in order to avoid biases in the recovered parameters.

Two main families of models have been developed, both being able to incorporate HMs.
One family relies on the Effective One Body (EOB) formalism 
\cite{Buonanno:2000ef, Damour:2009kr, Cotesta:2018fcv, Nagar:2020pcj, Chiaramello:2020ehz, Ossokine:2020kjp, Ramos-Buades:2023ehm, Nagar:2021gss}, 
which maps the complicated general relativistic binary system into a problem governed by an effective Hamiltonian. 
EOB models tend to be accurate but are quite costly to generate, since for each waveform one needs to solve the 
Hamiltonian equation of motion.
On the other hand, the phenomenological waveforms \cite{Khan:2015jqa,Pratten:2020ceb,Estelles:2020osj} are based on 
analytical expressions, making use of the post-Newtonian formalism to model the inspiral, and on fits to numerical
simulations to describe the intermediate and merger-ringdown regimes. They tend to be faster to evaluate than the EOB models.
Both families, EOB and phenomenological, need to be calibrated with numerical relativity waveforms, 
computed by directly solving the Einstein equations in discretized form. The calibration makes sure that a model 
retains its accuracy even close to merger, where approximate treatments such as the the post-Newtonian or EOB formalisms 
are no longer applicable by themselves.

Besides the standard families, surrogate waveform models have been developed with the aim of reproducing the 
output of a target model and of making feasible the usage of the underlying model.
A first class of surrogates is designed to closely reproduce numerical relativity (NR) waveforms \cite{Blackman:2015pia, Varma:2018mmi, Blackman:2017dfb, Blackman:2017pcm, Varma:2019csw, Williams:2019vub, Rifat:2019ltp} and, accordingly, it is trained using only NR waveforms as input. NR surrogates are very accurate but they tend to be very short, due to the nature of the NR waveforms employed for training. For this reason, they are often hybridized using an analytical expression for the early inspiral.
Besides targeting NR waveforms, a second class of surrogates have been developed to accelerate EOB models \cite{Field:2013cfa, Purrer:2014fza, Purrer:2015tud, Cotesta:2020qhw, Gadre:2022sed}, even including HMs. While traditional surrogate 
models build an empirical interpolant on the waveform space, a more recent approach relies on performing a regression 
using machine learning techniques \cite{Chua:2018woh, Khan:2020fso, Thomas:2022rmc}.

Among others, \cite{Schmidt:2020yuu} introduced a machine learning surrogate model, based on a dimensionality 
reduction scheme followed by a regression. The framework was later applied to the generation of frequency domain signals from Binary Neutron Star (BNS) systems \cite{Tissino:2022thn}.
In this work, we extend the model to HMs and we improve the accuracy 
of the regression by employing artificial neural networks (ANN). Our model marks a step towards the development of a faster, yet precise, waveform model, and will help enable the accurate analysis of next-generation detector data.

We train our model on the widely used approximant \texttt{SEOBNRv4HM} \cite{Cotesta:2018fcv} to target systems with mass ratio $q \in [1,10]$ and dimensionless spin components between $[-0.9, 0.9]$.
Our model is able to generate waveforms with a maximum length of $t = 2 \, \frac{M}{\textrm{M}_\odot} \textrm{s}$, which amounts to $t \simeq 4.06\times 10^5 M$ in geometrized units, and achieves $10^{-4}$ median faithfulness (with tails up to $10^{-2}$) when averaged across a wide range in parameter space.
Our numerical experiments show that our model offers a substantial speed-up with respect to the original model, matching 
the speed of the state-of-the-art surrogate models.

Our methodology distinguishes itself from previous approaches \cite{Chua:2018woh, Khan:2020fso, Thomas:2022rmc} in several ways. First of all, to predict the phase $\phi_{\ell m}$ of each mode, we use three separate models for predicting the basis coefficients of the principal component decomposition. This ``distribution of tasks'' allows for more flexibility, a significant reduction in the total number of model parameters and fewer training waveforms. Secondly, we improve on choosing the hyperparameters of the ANNs. Whereas previous approaches arrive at their configuration of hyperparameters using somewhat limited heuristics, we introduce a more rigorous method by using Bayesian optimization to tune hyperparameters.
Lastly, and more importantly, even though we have not incorporated precession, we show that our approach is viable for the large range of parameters ${q \times \chi_{1z} \times \chi_{2z} = [1,10]\times [-0.9,0.9]\times[-0.9,0.9]}$. Previous approaches either focused solely on the dominant mode of non-precessing spin-aligned waveforms \cite{Khan:2020fso, Chua:2018woh} or on precessing HM waveforms but with a limited range in mass ratio $q \in [1,2]$ and with $\chi_{2}=0$ \cite{Thomas:2022rmc}.

This paper is organized as follows. In Sec.~\ref{sec:model} we introduce the details of the model presented here, 
stressing the differences with the model in \cite{Schmidt:2020yuu}.
Sec.~\ref{sec:performance} is devoted to the validation of our model: we will motivate our choice of several 
hyperparameters and perform an accuracy and speed study.
In Sec.~\ref{sec:end}, we present some final remarks and highlight future perspectives.

\section{Building the model}
\label{sec:model}

A non-precessing BBH can be described by four {\it intrisic} parameters, which specify the two BH masses $m_1$ and $m_2$ 
and the z-components of the two dimensionless spins, $\chi_\text{1z}$ and $\chi_\text{2z}$.
Since the total mass $M = m_1 + m_2$ acts as a scaling parameter, when generating non-precessing BBH signals one  
only needs to consider the mass ratio $q = m_1/m_2 \geq 1$ together with the spins. 
We refer to the relevant parameters as $\boldsymbol{\vartheta} = (q, \chi_\text{1z}, \chi_\text{2z})$.
Besides the masses and spins, the gravitational wave emitted by the system depends also on 
luminosity distance to the source $d_L$, the inclination angle $\iota$ of the source, and the reference phase 
$\varphi_0$; these are the {\it extrinsic} parameters.

As is standard, we expand the angular dependence on $\iota, \varphi_0$ of the {\it complex} waveform $h(t)$ 
in terms of a sum of spin -2 spherical harmonics.
A GW is then parameterized\footnote{Such parameterization is particularly convenient as it separates 
the waveform dependence over intrinsic and extrinsic parameters.} as~\cite{Estelles:2021gvs}:
\begin{align} \label{eq:h_parametrization}
	&h(t; d_L,\iota,\varphi_0, m_1, m_2, \chi_\text{1z}, \chi_\text{2z}) = h_+ + i h_\times \nonumber \\
		&\qquad= \frac{G}{c^2} \frac{M}{d_L}\sum_{\ell = 2}^{\infty} \sum_{m = -\ell}^{\ell} \tensor[^{-2}]{Y}{_{\ell m}}(\iota, \varphi_0) h_{\ell m}(t/M; \boldsymbol{\vartheta}),
\end{align}
where we refer to the functions $h_{\ell m}(t; \boldsymbol{\vartheta})$ as {\it modes} of the waveform. We note that that, 
for non-precessing systems, $h_{\ell m} = (-1)^\ell h^*_{\ell -m}$, hence we will only consider modes with $m>0$.

The mode $(\ell, m) = (2,2)$ is the largest in amplitude, hence it is often referred to as the {\it dominant mode}. 
The other sub-dominant modes are usually few orders of magnitude smaller in amplitude and become more relevant 
(and measurable) for high mass ratios \cite{Mills:2020thr, LIGOScientific:2020stg, LIGOScientific:2020zkf}.

In this work, we introduce a machine learning model to perform a regression
\begin{align}\label{eq:objective}
	(q, \chi_\text{1z}, \chi_\text{2z}) &\longmapsto h_{\ell m}(t; \boldsymbol{\vartheta})
\end{align}
for each mode $(\ell,m)$.
The regression is designed to reproduce waveforms from a given data set; such waveforms can be generated 
by {\it any} time-domain approximant.

We decompose each mode in an amplitude term $A_{\ell m}$ and a phase term $\phi_{\ell m}$ as follows:
\begin{equation}\label{eq:amp_ph_decomposition}
	h_{\ell m}(t; \boldsymbol{\vartheta}) = A_{\ell m}(t; \boldsymbol{\vartheta}) e^{i \phi_{\ell m}(t; \boldsymbol{\vartheta})},
\end{equation}
and, for each mode, we perform a regression for amplitude and phase separately. The regression scheme closely follows \cite{Schmidt:2020yuu} and relies on:
\begin{enumerate}[label=(\alph*)]
	\item A suitable vector representation of the regression target by choosing a fixed time grid;
	\item A principal component analysis (PCA) model to reduce the dimensionality of each waveform;
	\item An artificial neural network (ANN) regression to learn the dependence on $\boldsymbol{\vartheta}$ of the 
	reduced waveform.
\end{enumerate}

While the first two elements are unchanged from the previous work, the ANN regression is first introduced here. Indeed a NN has more representation power than the Mixture of Experts (MoE) model \cite{Jacobs1991AdaptiveMoE}, used in \cite{Schmidt:2020yuu}: the change was needed to achieve better accuracy for the model.

\subsection{Data set creation}
\label{sec:dataset_creation}

To construct a data set, we follow \cite{Schmidt:2020yuu} and we set a dimensionless time grid. We construct the grid 
by setting $D$ points equally spaced in $\tau^\alpha$, where $\tau$ is the physical time scaled by the total mass of 
the system $M$: $\tau = t/M$. Using the findings of \cite{Schmidt:2020yuu}, we set $D = \text{2000}$ and 
$\alpha = \text{0.5}$.
This is a good compromise between the need of having a faithful representation of the waveform (which requires a large grid) and the need of having a compact model (which points to a sparse grid).
The waveforms are time-shifted so that the peak of the amplitude of the $(2,2)$ mode happens at $\tau=0$. The grid starts at (scaled) time $\tau_\text{min} = -\tau_0$, where $\tau_0$ sets the length of the waveform that our model is able to generate (as a function of the total mass $M$).
We choose $\tau_0 = 2 s/\textrm{M}_\odot$, which in geometrized units (i.e. with $G=c=1$) amounts to a waveform length of $t \simeq 4.06\times 10^5 M$.
We populate the data set with $68000$ waveforms.

To make sure that the distribution of $q$ is skewed towards towards the boundaries, where the regression is less accurate, we sample the mass ratio $q$ in the range $[1,10]$ with the following procedure:
\begin{itemize}
	\item We sample $q_1, \hdots, q_5 \sim \mathcal{U}_{[1,10]}$;
	\item We sample $x \sim \mathcal{U}_{[0,1]}$;
	\item We select $q$, based on the value of $x$:
	\begin{itemize}
		\item If $x \in [0,0.3)$, $\min q_1, \hdots, q_5$,
		\item If $x \in [0.3, 0.8)$, $q_1$,
		\item If $x \in [0.8, 1]$, $\max q_1, \hdots, q_5$,
	\end{itemize}
\end{itemize}
where $\mathcal{U}_{[a,b]}$ is the uniform distribution in $[a,b]$.
The spins are drawn uniformly in the range $[-0.9, 0.9]$.

Once a time grid is set, we evaluate all the modes (amplitude and phase) on the time grid and represent them as vectors in $\R^D$.
We then create a data set $\{X, Y\}$ of $N$ elements. Each row of the data set is of the form:
\begin{align}
	X &= [q, \chi_\text{1z}, \chi_\text{2z}] \\
	Y &= [\boldsymbol{A}^T_{\ell m}, \boldsymbol{\phi}^T_{\ell m}, \hdots ] 
\end{align}
The data set $Y$ gathers the amplitude and phase for the different modes in the data set.
We include all the modes available in \texttt{SEOBNRv4HM}: $(\ell, m) = \{(2,2),(2,1), (3,3), (4,4), (5,5)\}$.
In what follows we will refer to any of the vectors $\boldsymbol{A}_{\ell m}$ or $\boldsymbol{\phi}_{\ell m}$ as $\boldsymbol{f}$.
Note that we use the same grid for all the modes.

It is well known \cite{Blackman:2015pia, Varma:2018mmi} that in the case of a symmetric BBH, where $m_1\simeq m_2$ and $\chi_\text{1z} \simeq \chi_\text{2z}$, the amplitude of the odd-m modes vanishes, making the phase Eq.~\eqref{eq:amp_ph_decomposition} ill-defined.
Clearly this is a challenge for the regression, which is likely to perform poorly in such situations.
While several alternatives are available in the literature, we will address the challenge in future work. For the moment we will content ourselfs with a poor performance of the regression for the odd-m modes in the $m_1\simeq m_2$ region, as we will show in Fig.~\ref{fig:accuracy_hist} and Fig.~\ref{fig:countour_plots_modes}. This will have little impact for the overall faithfulness, due to the small amplitude of such modes.

\subsection{Dimensionality reduction}
\label{sec:PCA}

It is unfeasibile to perform a regression targeting a high-dimensional vector such as $\boldsymbol{f} \in \R^D$. 
For this reason, in \cite{Schmidt:2020yuu} we introduced a principal component analysis (PCA) dimensionality 
reduction scheme.
It is an {\it approximately} invertible linear mapping between a vector $\boldsymbol{f} \in \R^D$ in a large dimensional space to lower dimensional vector  $\boldsymbol{g} \in \R^K$:
\begin{align}
	\mathbf{g} = H (\mathbf{f} - \boldsymbol{\mu}) \label{eq:PCA_reduction_model},\\
	\hat{\mathbf{f}} = H^T \mathbf{g} + \boldsymbol{\mu}, \label{eq:PCA_reconstruction_model}
\end{align}
where $\boldsymbol{\mu} \in \R^D$ and $H$ is a $K \times D$ matrix.
The rows $H_{i:}$ of $H$, also called {\it principal components} (PC), form an orthonormal set of vectors, 
i.e.~${\sum_{k=1}^D H_{ik} H_{kj} = \delta_{ij}}$.
The PCs are the first $K$ eigenvectors of the $D \times D$ covariance matrix of the data set, as described in 
\cite[Sec. 12]{murphy2012machine}.

The mapping is only approximately invertible, in the sense that $\hat{\mathbf{f}}$ is only an approximation of the high dimensional vector $\mathbf{f}$. The quality of the approximation is controlled by the number $K$ of PCs considered: the more PCs, the more accurate the reconstruction of $\mathbf{f}$ is.

One can have a deeper insight on PCA considering the following formula for the reconstructed vector $\hat{\mathbf{f}}$ (setting $\boldsymbol{\mu}=0$ without loss of generality):
\begin{equation} \label{eq:perturbative_exp}
	\hat{\mathbf{f}} = \sum_{i=0}^{K-1} \langle \mathbf{f} | H_{i:} \rangle \; H_{i:}
\end{equation}
where $\langle \mathbf{a} | \mathbf{b} \rangle = \sum_{i=0}^{D-1} a_i b_i$ is the {\it Euclidean} scalar product between two vectors $\mathbf{a}, \mathbf{b} \in \R^D$.
Since less important PCs are more orthogonal to data, the typical magnitude of $g_i = \langle \mathbf{f} | H_{i:} \rangle$ decreases as $i$ increases.\footnote{For this reason PCA can be seen as a perturbative expansion on the basis vectors $H_{i:}$, where the accuracy is roughly measured by the eigenvalues of the first neglected PC. Increasing the number K of PCs considered increases the accuracy of the model (but also the complexity of the model).}.
As a consequence, the regression for a lower order PC needs to be more accurate than the one for the higher order PC. This will be taken care of by a suitable choice for the loss function for the regression (see next section).

Following \cite{Schmidt:2020yuu}, in this work we employ for each mode six PCA compontents for the phase model and four for the amplitude.
While it is plausible that an optimal number of components may vary for different modes, we opt for simplicity by employing the same number of PCA components for all modes, a configuration tuned based on the $(2,2)$ mode only.

\begin{figure*}[t]
	\begin{tabular}{cc}
		\includegraphics[scale = 0.9]{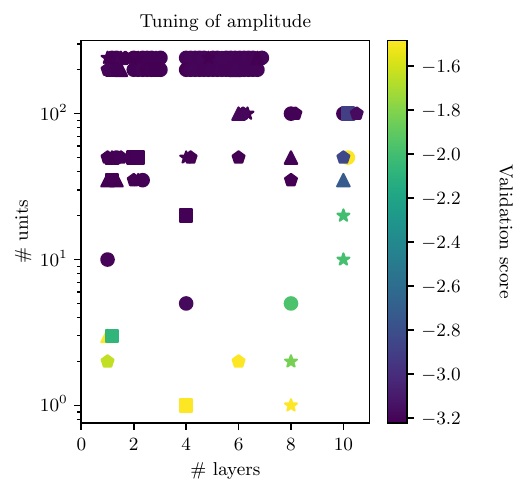} &
		\includegraphics[scale = 0.9]{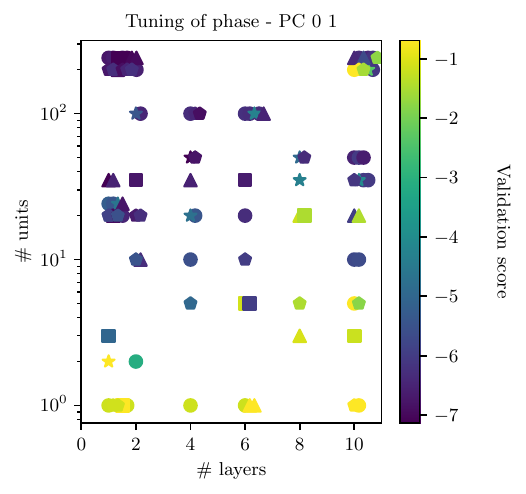} \\
		\includegraphics[scale = 0.9]{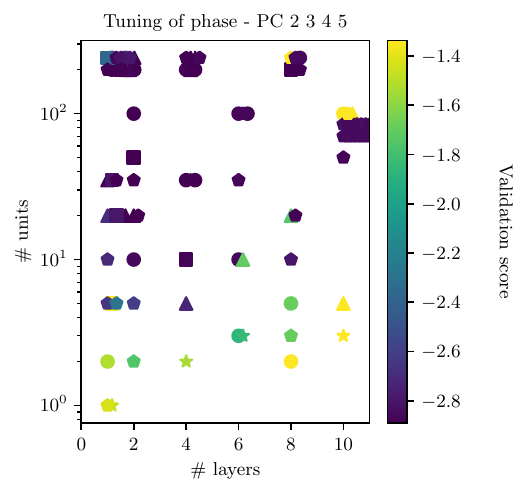} &
		\includegraphics[scale = 0.9]{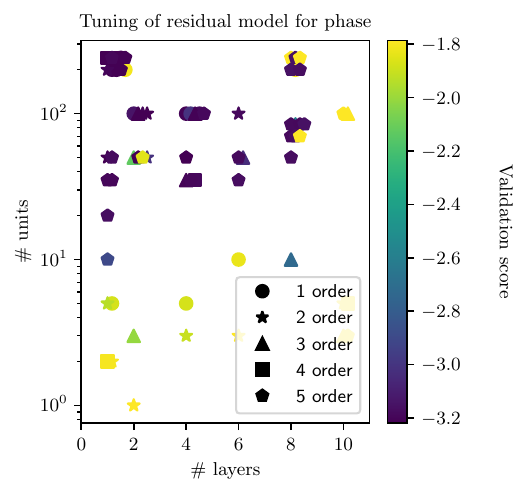}
	\end{tabular}
	\caption{Results from the validation of our ANN models, using the $\ell, m = 2,2$ mode data set. We tune the number of layers and the number of features per layer, together with the features and the polynomial order for the data augmentation.
	Each panel in the figure refers to a different ANN, taking care of different parts of the regression, as described in Sec.~\ref{sec:NN}.
	For each regression, we train 100 ANNs with different choices of hyperparameters. Each point in the plot, refers to a trained network and it is colored with the logarithm of the loss function computed on the validation data, referred to as the {\it validation score}. Note that we do not report the features used for data augmentation, so that the plot is degenarate in this quantity.
	}
	\label{fig:tuning}
\end{figure*}

\subsection{Neural network regression}
\label{sec:NN}

An artificial neural network (ANN) is a popular regression model, consisting of a powerful parametric function, 
whose parameters (or weights), when properly set, can represent a large variety of relations between input and output \cite{Bishop2006machinelearning, murphy2012machine, Goodfellow-et-al-2016}.
An ANN is built by stacking together $N_\text{L}$ layers in such a way that the output of a layer is the input of the 
following layer. Each layer is a function $L: \R^{D^\prime} \rightarrow \R^{D^{\prime\prime}} $ and has the following 
functional form:
\begin{equation}
	\boldsymbol{y} = a(W^{\prime}_{\prime\prime}\boldsymbol{x})
\end{equation}
Where $W^{\prime}_{\prime\prime}$ is a $D^{\prime\prime} \times D^{\prime}$ matrix and $a: \R \rightarrow \R$ is an activation function that acts element-wise on the vector $W^{\prime}_{\prime\prime}\boldsymbol{x}$.
Each component $y_i$ of the output of the layer is called a node and the number of nodes is a tunable parameter, controlling the representative power of the layer.

An ANN $\mathcal{N}$ is obtained by composing $N_\text{L}$ different layers (each with a suitable number of nodes):
\begin{equation}
	\mathcal{N}_W = L_{N_\text{L}} \circ \hdots \circ L_2 \circ L_1
\end{equation}
where we denote by $W$ the set of all the parameters the ANN depends on.

The number of layers, together with the number of nodes per layer are hyperparameters that need to be carefully chosen, to balance model accuracy and model complexity.
Another important choice is the activation function: several possible choices are possible, the most popular being the {\it sigmoid}, the hyperbolic tangent or the so called ReLU function. In our work, we consider the sigmoid function between all layers, except for the very last layer which has linear/identity activation so that negative values are also possible.

Once the ANN is set up, we need to set its weights to the values that achieve our regression task.
This procedure is called training, where we minimize a loss function with respect to the weights $\boldsymbol{W}$ of the model.
The loss function depends on the data set at hand ${\{\boldsymbol{x}_i, \boldsymbol{y}_i\}}$.
Mathematically, the weights are given by:
\begin{equation}\label{eq:loss_general}
	\boldsymbol{W} = \argmin_W \mathcal{L}(W; \{\boldsymbol{x}_i, \boldsymbol{y}_i\}_i)
\end{equation}

The minimization of the loss function is performed by stochastic gradient descent (SGD), as implemented by the \texttt{Nadam} algorithm \cite{dozat.2016}, which combines the popular \texttt{Adam} algorithm \cite{Kingma2014AdamAM} with the Nesterov momentum. The optimization relies on the gradients $\partial_W \mathcal{L}$ of the loss function, computed through the 
back-propagation algorithm \cite{backprop}.

To perform our regression $\theta \longmapsto \boldsymbol{g}$, we employ an ensemble of networks that suitably combined delivers accurate results.
To improve the representative power or the ANN, we employ feature augmentation on the vector $\boldsymbol{\vartheta} = (q, \chi_\text{1z}, \chi_\text{2z})$, effectively using the augmented vector $\tilde{\boldsymbol{\vartheta}}$ as input for the regression. Although different ANNs will need different features, we will for convenience 
abuse the notation $\tilde{\boldsymbol{\vartheta}}$ to denote any augmented vector. Indeed, the features to add need to be chosen with a validation process: this will be discussed in the next section.

Before the training, the regression targets $\boldsymbol{y}_i$ are scaled such that $\boldsymbol{y}_i \rightarrow \frac{\boldsymbol{y}_i}{\boldsymbol{w}}$, where $\boldsymbol{w}$ keeps the maximum of $|\boldsymbol{y}_i|$ along each axis. 
In this way all the regression targets span the same order or magnitude, facilitating the ``learning" task.

For the amplitude $\boldsymbol{A}_{\ell m}$ of each mode, we employ a single ANN $\mathcal{N}_{A_{\ell m}}$ that predicts the first four PCA components.
The predicted amplitude $\hat{\boldsymbol{A}}_{\ell m}$, including the PCA reconstruction, has the following form:
\begin{equation}\label{eq:amp_pred}
	\hat{\boldsymbol{A}}_{\ell m}(\boldsymbol{\vartheta}) = \boldsymbol{\mu}_{A_{\ell m}} + H_{A_{\ell m}}^T \mathcal{N}_{A_{\ell m}}(\tilde{\boldsymbol{\vartheta}}).
\end{equation}

For the phase $\boldsymbol{\phi}_{\ell m}$, we employ one ANN $\mathcal{N}_{\phi_{\ell m}\text{- 01}}$ to predict only the first two PCA components.
Another ANN will take care of the remaining components $\mathcal{N}_{\phi_{\ell m}\text{- 2345}}$.
On top of this, we build an additional ANN $\mathcal{N}_{\phi_{\ell m}\text{- residual}}$ to target the residual of the predictions of $\mathcal{N}_{\phi_{\ell m}\text{- 01}}$.
The scheme makes sure that the first two PCs are predicted with much larger accuracy than the others. Indeed, the reconstructed WF depends largely on the first two components and a small fractional error can potentially have a large impact on the overall accuracy.

The predicted phase $\hat{\boldsymbol{\phi}}_{\ell m}$ is then given by
\begin{align}
	\hat{\boldsymbol{\phi}}_{\ell m}(\boldsymbol{\vartheta}) &= \boldsymbol{\mu}_{\phi_{\ell m}}  + \nonumber\\
	& H_{\phi_{\ell m}}^T 
	\begin{pmatrix}
        \mathcal{N}_{\phi_{\ell m}\text{- 01}}(\tilde{\boldsymbol{\vartheta}}) + \mathcal{N}_{\phi_{\ell m}\text{- residual}}(\tilde{\boldsymbol{\vartheta}}) \\
        \mathcal{N}_{\phi_{\ell m}\text{- 2345}}(\tilde{\boldsymbol{\vartheta}}) \hfill
	 \end{pmatrix} \label{eq:ph_pred}
\end{align}
We train our model using the PCA data set, obtained by PCA reducing the training set. Each ANN is trained using the following loss function:
\begin{equation}\label{eq:loss}
	\mathcal{L} = \frac{1}{N} \sum_{i=1}^N \Bigl((\mathcal{N}(\boldsymbol{\vartheta}_i) - \boldsymbol{y}_i)\Bigr)^2 \; \boldsymbol{w},
\end{equation}
where $\boldsymbol{y}_i$ is the (scaled) regression target of each network and $\boldsymbol{w} \in \R^K$ takes into account the fact that different PCs have different orders of magnitude.

The network is implemented and trained using the python package \texttt{keras} \cite{chollet2015keras}, built on \texttt{tensorflow} backend \cite{tensorflow2015-whitepaper}.

\section{Performance study}
\label{sec:performance}
In this section, we first study how the model performance depends on the different choices of hyperparameters 
(network architecture, learning rate, features, ...).
The architecture details of the model (chosen after hyperparameters tuning) are reported in Tab.~\ref{tab:model}.
We then evaluate the faithfulness of our model and report the speed up that we obtain when using our surrogate instead of the training model \texttt{SEOBNRv4HM}.
In what follows, we will refer to our model as \texttt{mlgw-SEOBNRv4HM}.

To measure the discrepancy between two waveforms  $h_1$, $h_2$, we define a scalar product
\begin{equation}
	\rescalar{h_1}{h_2} = 4 \Re \int_{0}^{\infty} \text{d}f \; \frac{{\tilde{h}^*}_1(f) \tilde{h}_2(f)}{S_n(f)},
\end{equation}
where $\tilde{\phantom{h}}$ denotes the Fourier transform and $S_n(f)$ is the power spectral density (PSD) of the detector's noise.
We can use the scalar product to arrive at a normalized waveform, $\hat{h} = \frac{h}{\sqrt{\rescalar{h}{h}}}$.

To measure the discrepancy between two individual modes $h^1_{\ell m}$ and $h^2_{\ell m}$, we define the {\it match} 
$\mathcal{M}$:
\begin{equation}\label{eq:match}
	\mathcal{M} = \max_{t,\phi} \; \rescalar{\hat{h}^1_{\ell m}}{\hat{h}^2_{\ell m} e^{i 2\pi ft + i \varphi}}
\end{equation}
where $h e^{i 2\pi ft + i \varphi}$ denotes (with a slight abuse of notation) $h$ translated in time by a factor of $t$ and with its phase shifted by $\varphi$.
We call {\it mismatch} the quantity $\mathcal{F} = 1 - \mathcal{M}$.

The match defined above amounts to the search statistics being used for matched filtering searches of non-precessing/non-HM signals \cite{Harry:2016ijz}.
A different statistic is needed to search for HM signals, hence the match defined above is not suitable to compare two different waveforms with HM content 
as in Eq.~\eqref{eq:h_parametrization}.
In this case, we need to compare the two polarizations $h_+$, $h_\times$ of a waveform with a signal $s$ observed at the 
detector:
\begin{equation}
	s = F_+ h_+ + F_\times h_\times,
\end{equation}
where $F_+, F_\times$ are called antenna pattern functions, depending on the sky location of the source and on the polarization angle \cite{Maggiore:2007ulw}.

Following \cite{Harry:2017weg}, we introduce the 
{\it symphony match} 
between a signal $s$ and a waveform $h$:
\begin{equation}\label{eq:match_sym}
\mathcal{M}_\mathrm{sym} 
	= \max_t \;
		\frac{ \rescalar{\hat{s}}{\hat{h}_+}^2 + \rescalar{\hat{s}}{\hat{h}_\times}^2 - 2 \rescalar{\hat{h}_\times}{\hat{h}_+} \rescalar{\hat{s}}{\hat{h}_+} \rescalar{\hat{s}}{\hat{h}_\times}}
		{1-\rescalar{\hat{h}_\times}{\hat{h}_+}^2}.
\end{equation}
Note that $\mathcal{M}_\mathrm{sym}$ depends on the signal $s$, hence it depends on the sky location and polarization angle.
As above, we define the symphony mismatch as $\mathcal{F}_\mathrm{sym} = 1 - \mathcal{M}_\mathrm{sym}$.

In what follows, we always use a constant (i.e.~flat) PSD. While this certainly does not correspond to any actual detector, it makes sure that all the frequencies are weighted equally, hence giving a detector agnostic measure of the mismatch.

\subsection{Hyperparameter tuning}
\label{sec:hyperparameter}

\begin{table}[t]
	\begin{tabular}{ l|c|c|c|c } 
		Network & \texttt{n-layers} & \texttt{units} & \texttt{features} & \texttt{order} \\
		\hline\hline
			$\mathcal{N}_{A_{\ell m}}$
				& 1 & 35 & $\mathcal{M}_c, \chi_\text{eff}$ & 1 \\
			$\mathcal{N}_{\phi_{\ell m}\text{- 01}}$ 
				& 2 & 50 & $\mathcal{M}_c, \eta, \log q, \chi_\text{eff}$ & 3 \\
			$\mathcal{N}_{\phi_{\ell m}\text{- 2345}}$
				& 1 & 50 & $\mathcal{M}_c, \eta, \log q, \chi_\text{eff}$ & 1 \\
			$\mathcal{N}_{\phi_{\ell m}\text{- residual}}$
				& 5 & 50 & $\mathcal{M}_c, \eta, \log q, \chi_\text{eff}$ & 2 \\
	\end{tabular}
	\label{tab:model}
	\caption{Architecture of the 4 ANNs employed to generate each mode. For each ANN we report the number of layers and the number of units per layer. 
	We perform data augmentation by adding all the polynomials terms in the chosen features.
	The architecture has been chosen after hyperparameter tuning (see Fig.~\ref{fig:tuning}).
	Among other features, we use the chirp mass $\mathcal{M}_c =\frac{(m_1 m_2)^{3/5}}{(m_1+m_2)^{1/5}}$, the symmetric mass ratio $\eta = \frac{m_1 m_2}{(m_1+m_2)^2}$ and the effective spin parameter $\chi_\text{eff} = \frac{m_1 \chi_\text{1z} +m_2 \chi_\text{2z}}{m_1+m_2}$
	}
\end{table}

\begin{figure}[t]
	\centering
	\includegraphics[scale = 1]{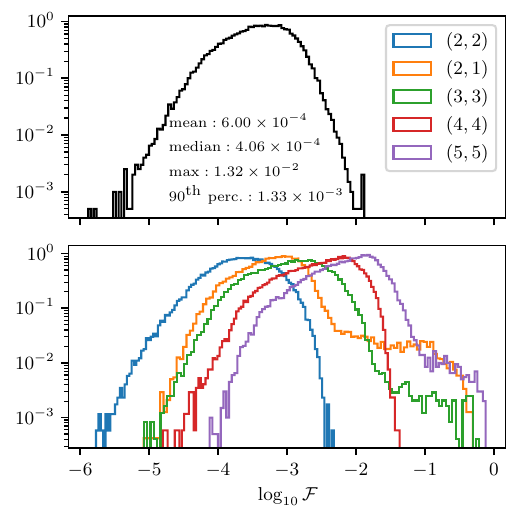}
	\caption{We report the results of the mismatch between the $50000$ test waveforms produced by \texttt{mlgw-SEOBNRv4HM} and by the training model \texttt{SEOBNRv4HM}.
	In the top panel, we report the histogram of the ``symphony" mismatch $F_\text{sym}$ for the overall waveforms, where we compare the $h_+$ and $h_\times$ polarizations (see Eq.~\eqref{eq:perturbative_exp}). For the computation, we set random sky location. We also report the median, the mean and the maximumum mismatch, together with the value of the $90^\text{th}$ percentile.
	In the bottom panel, we report the histograms for the mismatches computed mode by mode.
	The composition of the test set is described in the text.
	}
	\label{fig:accuracy_hist}
\end{figure}

\begin{figure*}[t]
	\centering
	\includegraphics[width=\textwidth]{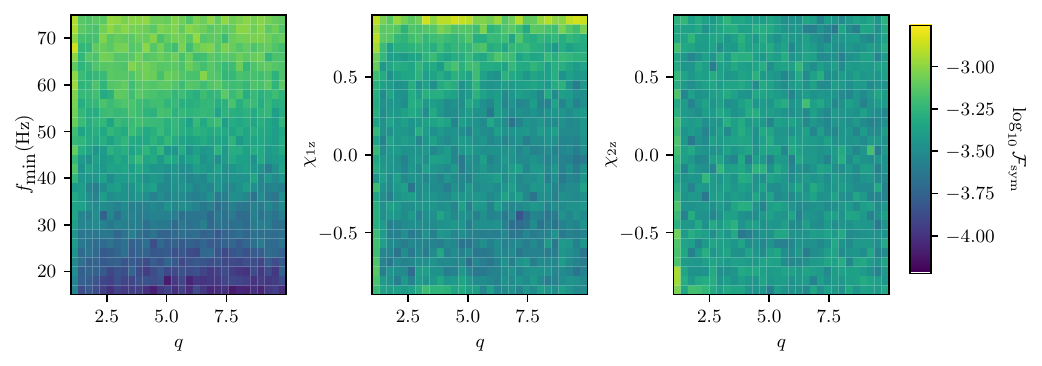}
	\caption{
	Depency of the ``symphony" mismatch $F_\text{sym}$ between \texttt{mlgw-SEOBNRv4HM} and the training model \texttt{SEOBNRv4HM}, as a function of some chosen waveform orbital parameters. The mismatch is computed on the $50000$ waveforms on the test set described in the text.
	On the left plot, we display the mass ratio and the starting frequency $q-f_\text{min}$ on the two axis, while we consider the the effect of spins on the center and right plot by showing the variables $q-\chi_\text{1z}$ and $q-\chi_\text{2z}$ respectively. 
	Each bin is colored according to the \textit{average} mismatch and the three plots shares the same color scale.
	We note that \texttt{mlgw-SEOBNRv4HM}'s faithfulness tends to decreases for low values of $q$, large positive values of $s_\text{1z}$ and higher values of $f_\text{min}$.
	}
	\label{fig:countour_plots}
\end{figure*}

\begin{figure*}[t]
	\centering
	\includegraphics[width=\textwidth]{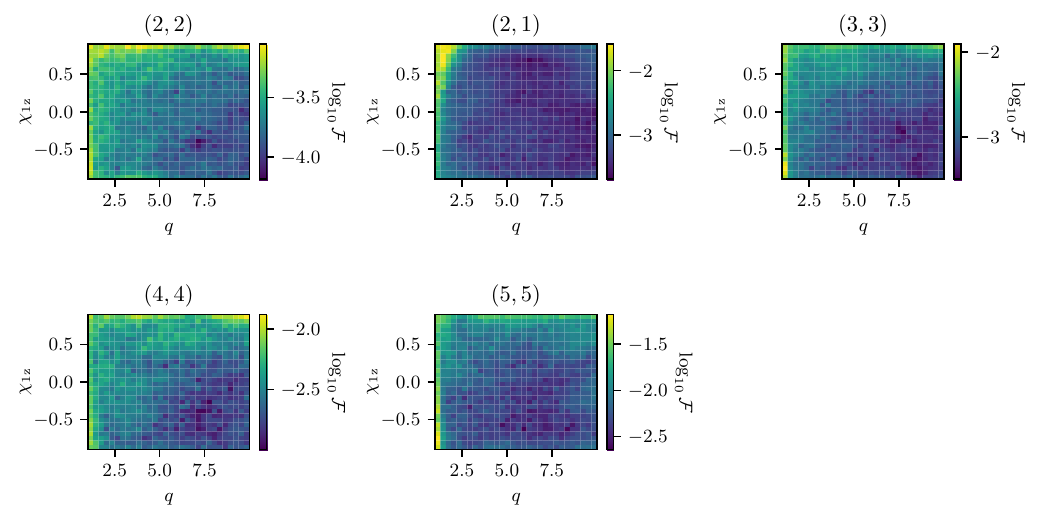}
	\caption{
	For each mode, we report the mismatch between \texttt{mlgw-SEOBNRv4HM} and the training model \texttt{SEOBNRv4HM}, as a function of $q$ and $s_\text{1z}$. The mismatch is computed on the $50000$ waveforms on the test set described in the text.
	Each bin is colored according to the \textit{average} mismatch.
	We note that the performance between different modes can vary significantly and in general they decrease for low values of $q$ and high values of spins.
	}
	\label{fig:countour_plots_modes}
\end{figure*}

The performance of the model depends on a number of crucial choices about some non-trainable parameters, usually called 
hyperparameters. The hyperparameters define the architecture of the ANN as well as some parameters relevant to the training.
Setting the right values for the hyperparameters is crucial for the ANN performance, as one needs to balance 
between accuracy and speed; this procedure is called {\it hyperparameter tuning} and can be done automatically to 
optimize manual work and to make sure to find a good minimum.

We optimize the following hyperparameters:
\begin{itemize}
	\item \texttt{n-layers}: number of hidden layers in the ANN;
	\item \texttt{units}: number of nodes per hidden layer;
	\item \texttt{features}: features to use for data augmentation;
	\item \texttt{order}: the data will be augmented with all the monomials of the chosen features up the given order.
\end{itemize}

For each of the 4 ANN useful to produce a single mode (see Eqs.~(\ref{eq:amp_pred}-\ref{eq:ph_pred})), we train a network for different combinations of hyperparameters. The figure of merit of each 
hyperparameter choice is the logarithm of the loss function (see Eq.~\eqref{eq:loss}) evaluated on the validation set.
For our experiments we only use the data set of the $(2,2)$ mode and we employ the package \texttt{keras-tuner} \cite{omalley2019kerastuner}. Specifically, we use the Bayesian optimization tuner which somewhat prioritizes searching around more promising configurations.

We report our results in Fig.~\ref{fig:tuning}, where each combination of hyperparameters tested is represented in the \texttt{n-layers}-\texttt{units} plane and colored by the validation score. 
We can see that all the four ANN share the same trend: the most effective way to improve regression accuracy is to increase the number of units as opposed to the number of layers.
The number of layers is far more important than extra features or the polynomial order for data augmentation.

Furthermore, we note that the regressions for the amplitude and for the high phase PCs 
(i.e.~components 2,3,4,5) can be performed with a smaller model, compared to the models for the first two PCs of the phase. 
This can be explained by the fact that most of the physical information is stored in the first two components of the phase, making this a harder regression problem.

In table Tab.~\ref{tab:model} we report the final hyperparameter choice we made for each of the networks. The architectures are the same across the different modes considered.

As discussed above, we note that models $\mathcal{N}_{A_{\ell m}}$ and $\mathcal{N}_{\phi_{\ell m}\text{- 2345}}$ are very simple, having only one layer and a small polynomial order, while the other ANNs have a more complicated architecture.
We note here that an accurate ANN for the residuals of the phase is crucial to obtain a good accuracy: indeed $\mathcal{N}_{\phi_{\ell m}\text{- residual}}$ is the most complex model we employ, meaning that the residual phase data set is the ``hardest" to learn.

\subsection{Accuracy study}\label{sec:accuracy}

To test the accuracy of our model, we generate a test set with $50000$ randomly chosen waveforms generated with the training model \texttt{SEOBNRv4HM}.
The waveforms masses are characterized by a total mass $M = 20 \textrm{M}_\odot$ and by a mass ratio $q \in [1, 10]$, while the spins are chosen in the range $[-0.9, 0.9]$. The inclination angle $\iota$ and reference phase $\varphi_0$ are drawn uniformly from a sphere.
To vary the length in time of the waveforms considered, we sample the starting frequency $f_\textrm{min}$ uniformly in the range $[15, 75]$ Hz.

In Fig.~\ref{fig:accuracy_hist}, we report the histogram of the distribution of the mismatches between \texttt{mlgw-SEOBNRv4HM} and the test waveforms. The upper part refers to the mismatches Eq.~\eqref{eq:match_sym} computed on the overall waveforms (with sky location sampled uniformly over the sky); the lower box refers to mismatches computed mode by mode with Eq.~\eqref{eq:match}.

First of all, we note that the model shows very high faithfulness. With a median value of $4\times 10^{-4}$ and with virtually no signals with a 
``symphony" mismatch higher than $10^{-2}$, the accuracy of \texttt{mlgw-SEOBNRv4HM} matches the accuracy of other state-of-the-art surrogate models \cite{Cotesta:2020qhw, Khan:2020fso, Gadre:2022sed} and the accuracy of the training model \texttt{SEOBNRv4HM} in reproducing numerical relativity waveforms \cite{Cotesta:2018fcv}.
The faithfulness for the $(2,2)$ mode is even higher, with no signals with mismatch exceeding $2\times 10^{-3}$.
On the other hand, the higher order modes are less accurately reproduced than the dominant mode. In particular, for the modes $(2,1), (3,3), (5,5)$ a limited number of waveforms show very high mismatches $\mathcal{O}(1)$. See below for more discussion.

In Fig.~\ref{fig:countour_plots} we report the dependence of the ``symphony" mismatch as a function of the different orbital parameters.
From the figure, it is manifest that the model has very stable performance across the parameter space.
The faithfulness decreases for high positive values of the spin of the first object $s_\text{1z}$ and for mass ratio $q\sim 1$. Despite this, in such ``extreme" regions, the average mismatch is still of the order of $10^{-4}$.
The performance of the regression does not depend on $\chi_\text{2z}$, since the quantity plays a very little role in defining the waveform features.

Longer waveforms, characterized by a lower $f_\textrm{min}$, tends to show higher faithfulness. As longer waveforms are dominated by the inspiral phase, we can conclude that our model is more successful in reproducing the inspiral as opposed to the merger and the ringdown, prevalent in short waveforms.
This feature is very important for the extension of our model to longer time grids, a necessity for analyzing binary neutron star systems or for applications in next-generation detectors \cite{Purrer:2019jcp}.

In Fig.\ref{fig:countour_plots_modes}, for each mode we report the mismatch as a function of the mass ratio and of $s_\text{1z}$. One more time, we can see that the model faithfulness decreases for low mass ratios and for high spins. Moreover, the subdominant modes shows a poorer performance as compared to the dominant one: this was already observed in Fig.~\ref{fig:accuracy_hist}.

The observed decrease in faithfulness for sub-dominant modes needs some attention.
As discussed in Sec.~\ref{sec:dataset_creation}, symmetric systems with $q \simeq 1$ have a vanishing amplitude of the odd-m modes and a poorly defined phase and, as such, they correspond to ``outliers" in the data set. This clearly poses a challenge for both the PCA and the regression model for the amplitude, since modelling such a sharp feature of the data requires an enhanced model flexibility and more training examples.
This is consistent with the low performance of the fit at low $q$ observed in Fig.~\ref{fig:countour_plots_modes}.

This matter is well known and several mitigation strategies are available in the literature.
First of all, we might incorporate the vanishing behaviour of the amplitude in the functional model for the regression, as done in \cite{Blackman:2015pia}. Concretely, we could introduce a $q$-dependent amplitude scaling for the waveforms before adding them to the data set, resulting in a data set with amplitude time series of approximately the same magnitude.
Second, we might mitigate the effect of a poorly defined phase by transforming all the modes except the $(2,2)$ in the co-orbital frame \cite{Varma:2018mmi} according to:
\begin{equation}
	h_{\ell m} \rightarrow h_{\ell m} e^{-i\frac{m}{2}\phi_{22}}
\end{equation}
where $\phi_{22}$ is the phase of the $(2,2)$ mode.
Both the strategies above reduce the outlier nature of waveforms with $q \sim 1$ and will likely improve the quality of the fit.

Another straightforward alternative could deploy a larger network for such modes. Indeed, we tuned the hyperparameters on the $(2,2)$ mode (an ``easy" regression target). Performing a network tuning on the data set of HMs might reveal that our chosen architecture is not optimal.

Finally, we note that since for $q\sim 1$, the sub-dominant modes have a vanishing amplitude, a large mismatch in the 
sub-dominant mode for $q\sim 1$ has very little impact on the overall waveform Eq.~\eqref{eq:h_parametrization}, 
as shown in Fig.~\ref{fig:accuracy_hist}.
This explains why the overall mismatch is low, despite high mismatch for the HMs in some edge cases.

\subsection{Timing study}
\label{sec:timing}

\begin{figure}[t]
	\centering
	\includegraphics[scale = 1]{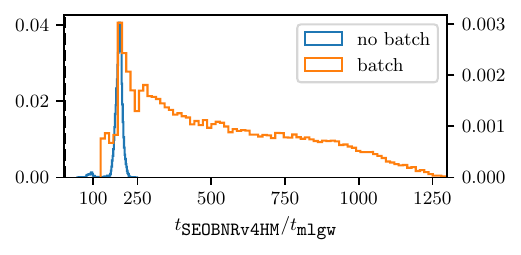}
	\caption{Speed up provided by \texttt{mlgw-SEOBNRv4HM} over the training model \texttt{SEOBNRv4HM}.
	In the histogram, we report the ration between the time $t_\texttt{SEOBNRv4HM}$ and the time $t_\texttt{mlgw}$ taken by the two models to generate each of the waveform in the test set.
	The two histograms are in different scales, reported on the left and right y axis for the ``no batch" and ``batch" case respectively.
	We note that \texttt{mlgw-SEOBNRv4HM} offers a speed up between a $150$ and $250$ with respect to the training model.
	\texttt{mlgw-SEOBNRv4HM} offers the option to generate waveforms in batches, effectively parallelizing some linear algebra operations. As shown in the plot, the batch generation provides an additional speed up, which can be as high as $1200$.
	}
	\label{fig:timing_hist}
\end{figure}

A speed-up in the waveform generation is the main motivation to build a ML waveform generator; for this reason it is crucial to assess the gain in waveform generation time.
For this reason, we use our test set to measure the ratio between the time to generate a waveform with \texttt{SEOBNRv4HM} and \texttt{mlgw-SEOBNRv4HM}. Our model offers further speed by generating waveform in batches: in this case, some operations are efficiently parallelized and happen more efficiently.
We report our findings in Fig.~\ref{fig:timing_hist}.

We achieve a speed-up ranging between a factor of $150$ and $250$, depending on the waveform characteristics. When waveforms are generated in batches of $100$, the speed-up can be substantially larger, reaching up to $1200$, although with considerable variance, mostly due to the waveform length.

The speed up achieved by \texttt{mlgw-SEOBNRv4HM} is comparable to the one obtained by the \texttt{SEOBNRv4HM\_ROM} surrogate model \cite{Cotesta:2020qhw}, which is the state-of-the-art frequency domain surrogate model trained on \texttt{SEOBNRv4HM}. \texttt{SEOBNRv4HM\_ROM} is obtained with standard techniques and it achieves a speed up ranging between 100 and 200.
The two results might not be directly comparable, since the comparison for \texttt{SEOBNRv4HM\_ROM} is performed in frequency domain and this involves computing the Fourier tranform of the \texttt{SEOBNRv4HM} waveform. As we perform the comparison in time domain, we omit the latter step, possibly obtaining lower values for the speed up as the ones obtained in \cite{Cotesta:2020qhw}. 

On the other hand, the speed up achieved by \texttt{mlgw-SEOBNRv4HM} is larger than the one obtained by the time domain surrogate model \texttt{SEOBNRv4PHMSur} \cite{Gadre:2022sed}. Indeed, the authors report a speed up always lower than $100$. However, also in this case, the comparison might be biased because the latter study also considers the effects of precession.

\section{Final remarks and future prospects}
\label{sec:end}

Building on our previous work \cite{Schmidt:2020yuu, Tissino:2022thn}, we generate a ML surrogate model \texttt{mlgw-SEOBNRv4HM} able to reproduce with very high fidelity the output of the widely used approximant \texttt{SEOBNRv4HM}. \texttt{mlgw-SEOBNRv4HM} can generate waveforms in a cuboid $q\times \chi_\text{1z} \times \chi_\text{2z} = [1,10]\times[-0.9,0.9]\times[-0.9,0.9]$ on a (reduced) time grid of maximum length of $2 \textrm{s}/\textrm{M}_\odot$, corresponding to waveforms of length $t \simeq 4.06\times 10^5 M$ in geometrized units.
Our model offers a two orders of magnitude speed up over the training model, without trading for accuracy, hence it is an attractive alternative for any data analysis application.
To encourage new applications, we release our code (and our trained model) publicly as a python package through the PyPI repository\footnote{The package is distributed under the name \texttt{mlgw} and is available at \url{https://pypi.org/project/mlgw/}.}.

Future work should also include precession. This can be achieved by means of the {\it spin twist} procedure \cite{Schmidt:2012rh, Schmidt:2014iyl, Pratten:2020ceb, Gamba:2021ydi}. It consist on a time dependent rotation of the plane of emission, resulting in a phase and amplitude modulation which approximates the effect of precession.
Training an ANN to predict the time dependent rotation is a promising step towards a complete ML surrogate model.

While the model is already applicable for most of the parameter estimation problems with current detectors, 
it is desirable to increase its range of validity, both in parameter space and in time span.
In principle, such an extension should be straightforward with the current network setup. On the other hand, due to an increased complexity of the regression task, probably more flexible architetures should be explored, using layers of different size. This would require a more careful (and computationally expensive) hyperparameters tuning.

An enhanced architecture should also benefit from sharing some parameters between models for different HMs - or even from treating the regression of the different modes as a large single regression problem. Indeed, the shapes of the different modes are correlated: for instance, the phases of two HMs are approximately proportional to each other. 
With the current architecture, the regression for each mode is carried on separately, hence each ANN needs to learn the WF behaviour independently. 
This could results in many redudant parameters in the network ensemble we introduced here.
Inserting parameter sharing inside the regression setup could result in a lighter ANN, which would lead to a reduced inference time.

Finally, we also strees that our PCA+ANN regression framework is fully general and in principle is applicable, with minimal modifications, to any chirp-like gravitational wave signal, such as extreme mass ratio inspirals (EMRI) \cite{Amaro-Seoane:2007osp, Amaro-Seoane:2012lgq} or BNS.
Extending the width of parameter space, enriching the BBH model with more physical effects and supporting a larger variety of systems will become mandatory for the next generation detectors \cite{Purrer:2019jcp, Owen:2023mid}, when fast and reliable waveform models will be needed to mitigate the huge computational cost posed by very long observed waveforms.
Our framework is ideal to achieve such an ambitious goal.


        \begin{acknowledgments}
          We thank Soumen Roy and Michael P{\"u}rrer for useful discussion and their precious comments.
          S.S. is supported by the research program of the Netherlands Organization for Scientific Research (NWO).
          This research has made use of data, software and/or web tools obtained 
          from the Gravitational Wave Open Science Center (https://www.gw-openscience.org), 
          a service of LIGO Laboratory, the LIGO Scientific Collaboration and the 
          Virgo Collaboration. LIGO is funded by the U.S. National Science Foundation. 
          Virgo is funded by the French Centre National de Recherche Scientifique (CNRS), 
          the Italian Istituto Nazionale della Fisica Nucleare (INFN) and the 
          Dutch Nikhef, with contributions by Polish and Hungarian institutes.
        \end{acknowledgments}

	

	\bibliography{biblio.bib}
	\bibliographystyle{ieeetr}

\end{document}